\documentclass[twocolumn,aps,prl,superscriptaddress]{revtex4}
\usepackage{graphicx}
\usepackage{amsmath}
\usepackage{amssymb}
\usepackage{braket}
\usepackage[colorlinks=true,linkcolor=blue,citecolor=blue,urlcolor=blue]{hyperref}

\begin{document}

\title{Efimov Enhanced Kondo Effect in Alkaline and Alkaline-Earth Atomic Gas Mixture}

\author{Juan Yao}
\affiliation{Institute for Advanced Study, Tsinghua University, Beijing, 100084,
China}

\author{Hui Zhai}
\affiliation{Institute for Advanced Study, Tsinghua University, Beijing, 100084,
China}
\affiliation{Collaborative Innovation Center of Quantum Matter, Beijing, 100084, China}

\author{Ren Zhang}
\email{renzhang@xjtu.edu.cn}
\affiliation{Department of Applied Physics, School of Science, Xi'an Jiaotong University, Xi'an, 710049, China}

\date{\today}
\begin{abstract}
Recent experiment has observed Feshbach resonances between alkaline and alkaline-earth atoms. These Feshbach resonances are insensitive to the nuclear spin of alkaline-earth atoms. Ultilizing this feature, we propose to take this system as a candidate to perform quantum simulation of the Kondo effect. An alkaline atom can form a molecule with an alkaline-earth atom with different nuclear spins, which plays the role of spin-exchange scattering responsible for the Kondo effect. Furthermore, we point out that the existence of three-body bound state and atom-molecule resonance due to the Efimov effect can enhance this spin-exchange scattering, and therefore enhance the Kondo effect. We discuss this mechanism first with a three-body problem in free space, and then demonstrate that the same mechanism still holds when the alkaline atom is localized by an external trap and becomes an impurity embedded in the alkaline-earth atomic gases.   

\end{abstract}
\maketitle

Quantum simulation with ultracold atomic gases has achieved great success in the past years, such as simulating the Bose and Fermi Hubbard models,  topological phase and the Kosterlize-Thouless transition \cite{BlochReview, EsslingerReviewFHM, CooperReviewTopo}.   
These quantum simulations certainly not only just recover known results in condensed matter counterpart, but also reveal lots of new universal physics by entering new parameter regimes or new circumstances such as highly non-equilibrium situations. The Kondo physics originates from a magnetic impurity embedded in a Fermi sea with spin-exchange interaction between impurity and itinerant fermions, and it is one of the standard examples for understanding impurity physics and transport in the strongly correlated regime. Simulating Kondo physics with cold atom system can also offer great promise for discovering new physics, for instance, by tuning the spin-exchange interaction to be quite strong or by quenching to the Kondo regime to watch the dynamical formation of the screening cloud. However, despite of great interests and many proposals of simulating the Kondo physics with cold atoms \cite{stoof,duan,rey,carmi,nishida1,demler,kikoin,rey2,kawakami,nishida2,ren-kondo,YT-kondo,ren-kondo2}, this goal has not succeeded yet.

Several requirements have to be satisfied in order to simulate the Kondo effect with ultracold atomic gases. First of all, one needs a mixture of two kinds of atoms, or two different internal states of the same atom, which have quite different ac polarizations. With a laser field, one can then create a deep potential to localize one component as impurities, while keep a shallow potential for another component to remain itinerant. Secondly, there must exist a spin-exchange interaction between the itinerant and the impurity atoms. Finally, if this spin-exchange interaction is weak, the Kondo temperature is several order of magnitude smaller than the Fermi temperature and therefore is not accessible by current cold atom experiments. Hence, one needs to tune the spin-exchange interaction to be sufficiently strong. For instance, recent works have suggested using the confinement-induced-resonance to enhance such spin-exchange interaction between the ground and the clock states of alkaline-earth atoms \cite{ren-kondo,YT-kondo,ren-kondo2,blochexp}.

\begin{figure}
\begin{center}
\includegraphics[width=0.4\textwidth]{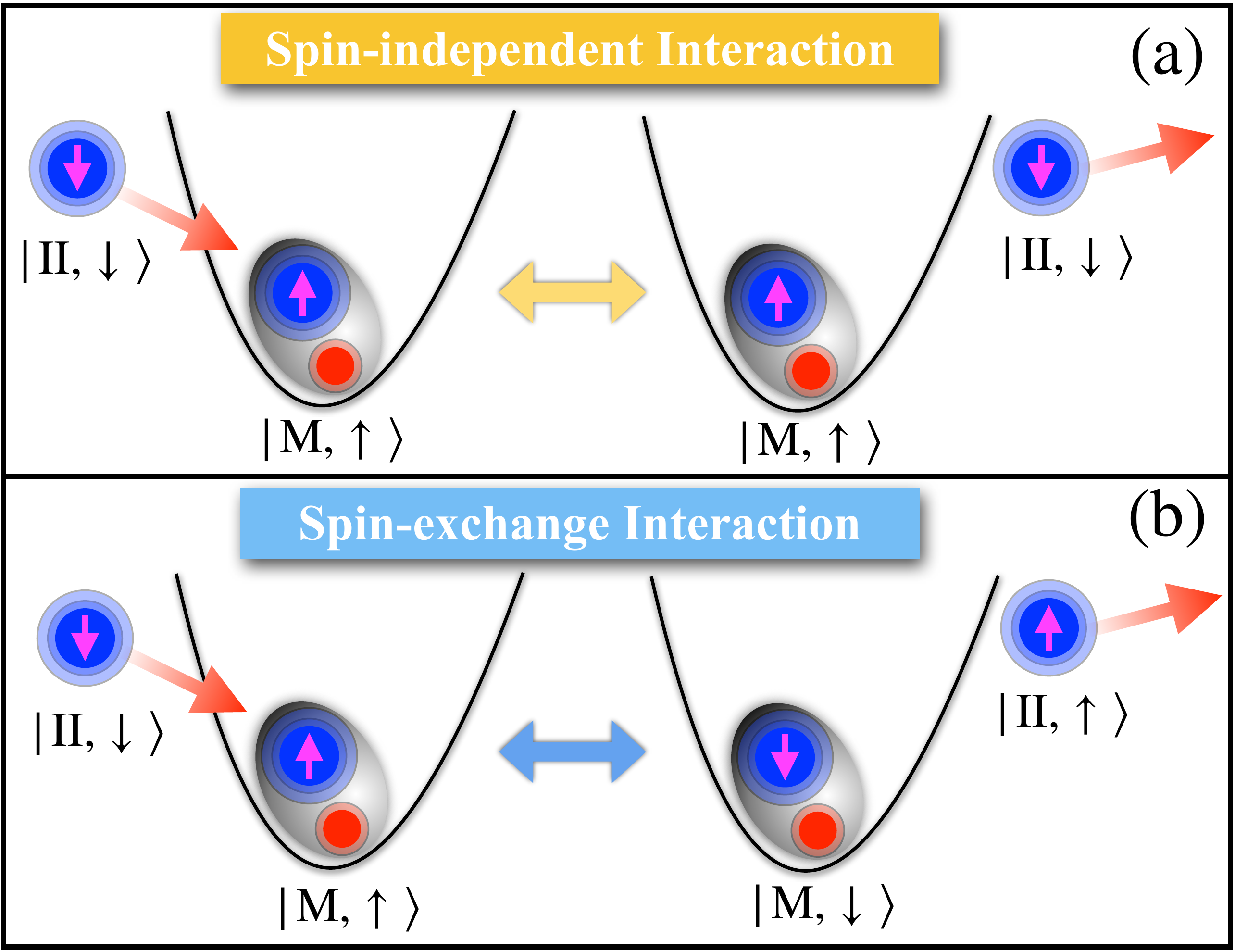}
\end{center}
\caption{(color online) Illustration of spin-independent (a) and spin-exchange (b) interaction in the collision between an AE atom and a two-body bound state composed by an AE and an A atom. The red and blue balls represent the A and the AE atoms, and denoted by {\rm I} and {\rm II} as short note of group-${\rm I}$ and group-${\rm II}$ atoms, respectively. $\uparrow$, $\downarrow$ denote two different nuclear spin states of the AE atoms. The bound state is denoted by M, standing for molecule, and it is spatially localized by external potential and plays the role of impurity. \label{schematic}}
\end{figure}

Recently an experiment from Amsterdam has successfully observed Feshbach resonances between alkaline (A) and fermionic alkaline-earth (AE) atoms \cite{AE-A Fesh}. The fermionic AE atoms have $N$ components due to the existence of nuclear spin. In this experiment with ${}^{87}$Rb and fermionic ${}^{87}$Sr, the shift of resonances positions between ${}^{87}$Rb and two different nuclear spin states of ${}^{87}$Sr is only about $~10$mG and the splitting cannot be resolved for a resonance with the width of loss feature about $300$mG. Thus here we take them as degenerate. The degeneracy of $N\geqslant 2$ Feshbach resonance does not require fine tuning of any parameters, and it is a novel feature in this system that does not exist in all previously observed Feshbach resonances. A natural question is whether we can use this feature to investigate new many-body physics. In this letter we propose that one can use these degenerate Feshbach resonances as a novel platform to simulate an SU($N$) Kondo physics, where the inelastic scattering between the AE atom and molecules with different nuclear spins plays the role of spin-exchange interaction. Moreover, we show that because the Efimov effect exists in this system, and the three-body bound state due to the Efimov effect can facilitate a resonant atom-molecule interaction, and hence, enhance the Kondo effect in this system.

{\it Spin-Exchange interaction and the Kondo Model.} The basic idea of this work is shown in Fig. \ref{schematic} by considering a three-body system. An A (group-{\rm I}) atom is labelled by $|{\rm I}\rangle$, and two AE (group-{\rm II}) atoms are labelled by $|{\rm II},s\rangle$, where for simplicity we only consider two out of $N$ nuclear spin components and they are denoted by $s=\uparrow,\downarrow$. Near the Feshbach resonance, the molecule formed by the A and the AE atom with spin-$s$ is labelled by $|\text{M},s\rangle$. Since we consider that the Feshbach resonance is independent of nuclear spin, the energies of $|\text{M},s\rangle$ with different $s$ are considered to be degenerate. In the molecular side of the resonance, the low-energy physics is dominated by the scattering between the remaining AE atom and the molecule. For instance, considering the scattering between an atom in $|{\rm II},\downarrow\rangle$ state and a molecule in $|\text{M},\uparrow\rangle$ state, the out-going state has two possibilities. For the first case the out-going state has the same internal state as the incoming state, as shown in Fig.\ref{schematic} (a), which is known as the elastic atom-molecule scattering. This gives rise to a spin independent interaction. For the second case the out-going state can be an atom in $|{\rm II},\uparrow\rangle$ state and a molecule in $|\text{M},\downarrow\rangle$ state, which is also known as the inelastic atom-molecule scattering. In this process, the nuclear spin of the AE atom is flipped. This gives rise to the spin-exchange processes.  

To be concrete, since the full Hamiltonian respects the symmetry of exchanging $\uparrow$ and $\downarrow$, so we should first introduce the bases that are either symmetric or anti-symmetric with respect to exchanging $\uparrow$ and $\downarrow$, denoted by $|\pm\rangle$ as 
\begin{equation}
\begin{aligned}
&|+\rangle=\frac{1}{\sqrt{2}} \left(|{\rm II}, \downarrow; {\rm M},\uparrow \rangle+ |{\rm II}, \uparrow; {\rm M},\downarrow\rangle\right),  \\
&|-\rangle=\frac{1}{\sqrt{2}} \left(|{\rm II}, \downarrow; {\rm M},\uparrow \rangle- |{\rm II}, \uparrow; {\rm M},\downarrow\rangle\right),
\end{aligned}
\end{equation}
and the atom-molecule interaction has to be diagonal in the bases, i.e. 
\begin{equation}
V_{{\rm II},\text{M}}({\bf r}={\bf r}_{\rm II}-{\bf r}_{{\text M}})=V_{+}({\bf r})|+\rangle\langle+|+V_{-}({\bf r})|-\rangle\langle-|,
\end{equation}
where ${\bf r}_{\rm II}$ and ${\bf r}_\text{M}$ denote the coordinates of the AE atom and the molecule, respectively. The atom-molecule scattering potential $V_{\pm}({\bf r})$ can be characterized by the Huang-Yang pseudo-potential as $V_{\pm}({\bf r})=\frac{2\pi\hbar^{2} a_{\pm}}{{\mu_{{\rm II},\text{M}}}}\delta({\bf r})\frac{\partial}{\partial r}\left(r\cdot\right)$, where $a_{\pm}$ is the atom-molecule scattering length and $\mu_{{\rm II},\text{M}}=m_{\rm II}m_{\rm M}/(m_{\rm II}+m_{\rm M})$ is the atom-molecule reduced mass. Here $m_{\rm M}=m_{\rm I}+m_{\rm II}$ is the mass of the molecule and $m_{\rm I}$, $m_{\rm II}$ are the masses of the A and the AE atom, respectively.

To connect this to the Kondo model, it is illuminating to write the bases back to 
\begin{align}
&V_{{\rm II},\text{M}}({\bf r})
=\frac{V_{+}({\bf r})+V_{-}({\bf r})}{2}\left(|{\rm II},\downarrow;{\rm M},\uparrow\rangle\langle {\rm II},\downarrow; {\rm M},\uparrow|+ \uparrow\leftrightarrow \downarrow \right)\nonumber\\
&+\frac{V_{+}({\bf r})-V_{-}({\bf r})}{2}\left({\rm II},\downarrow;|{\rm M},\uparrow\rangle\langle {\rm II},\uparrow ;{\rm M},\downarrow|+{\rm h.c.}\right),\label{spin-ex}
\end{align}
where the first term is the spin independent interaction (Fig. \ref{schematic} (a)) and the second term swaps the spins (Fig. \ref{schematic} (b)). The spin-exchange interaction is responsible for the Kondo effect. Because the A and AE atoms naturally have quite different ac polarization so it is easy to find a laser that creates a deep lattice to localize the A atom but a shallow lattice for the AE atoms such that they remain mobile. Since the A atom is localized, the molecule will be localized as well. When one expands the interaction Eq. \eqref{spin-ex} into a localized state for $|\text{M},s\rangle$ and in terms of momentum states for $|{\rm II},s\rangle$, it is straightforward to yield the Kondo model with the Kondo coupling depending on $a_{+}-a_{-}$ \cite{ren-kondo}.

\begin{figure}
\begin{centering}
\includegraphics[width=0.45\textwidth]{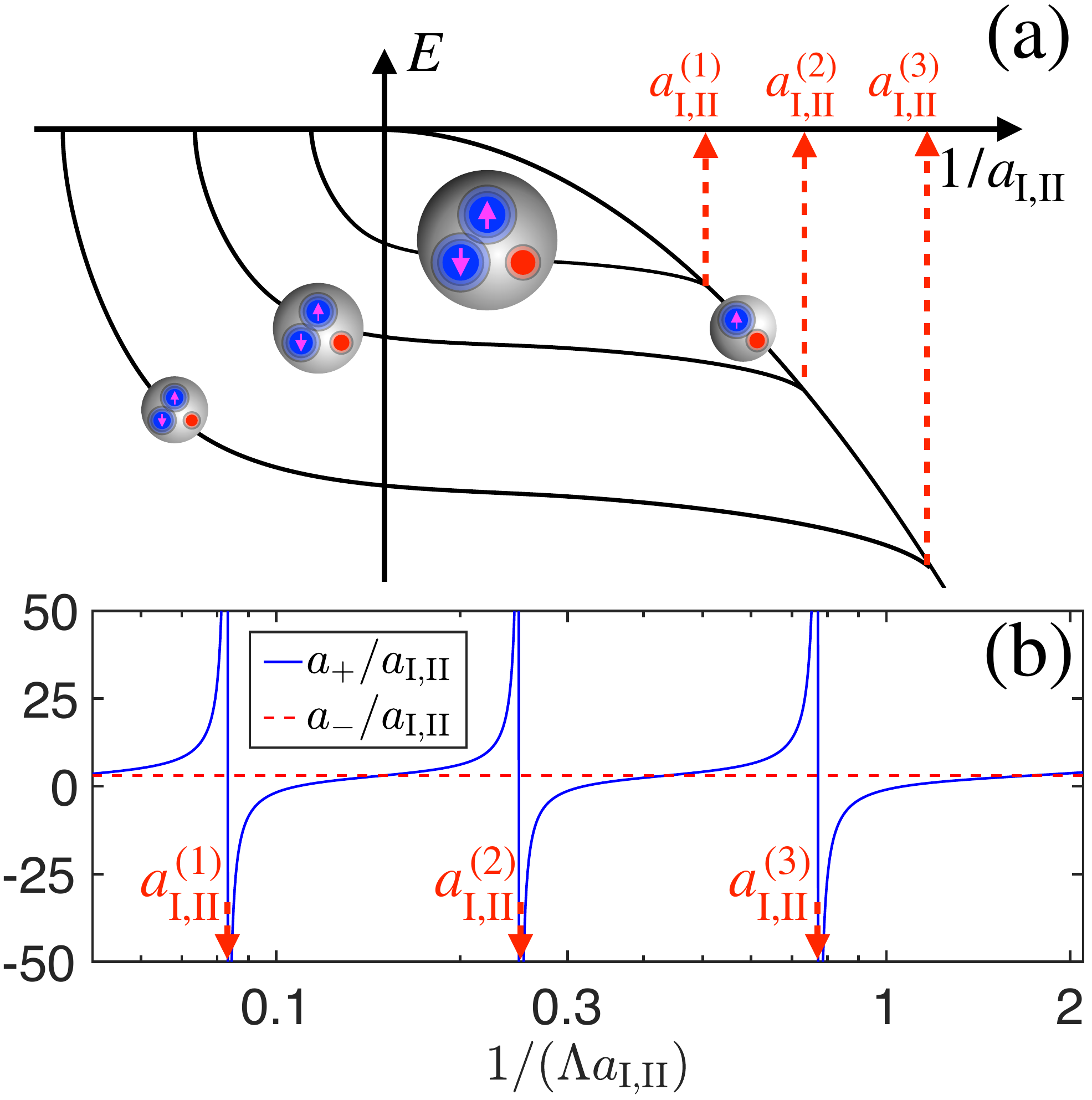}
\caption{(color online) 
(a): Typical molecule and trimer spectrum as a function of the atomic scattering length $a_{{\rm I, II}}$.  At the cross-point where the molecule and the Efimov trimer energies become degenerate, a resonance of the atom-dimer scattering length takes place. 
(b): The atom-dimer scattering length $a_{+}$ (blue solid curve) for the $|+\rangle$ channel shows multiple divergence due to the existence of Efimov states. While the scattering length $a_{-}$ (red dashed line) in $|-\rangle$ channel shows linear dependence on $a_{\rm I,II}$ with $a_{-}\approx3a_{\rm I,II}$. Here the mass ratio is fixed at $\delta=m_{\rm I}/m_{\rm II}=1/30$. }
\label{Fig2}
\end{centering}
\end{figure}

{\it Efimov Enhanced Spin-Exchange Interaction.} So far we have fulfilled the first two conditions for simulating the Kondo effect in this system, and the remaining task is to show how we can tune the Kondo coupling to be strong enough. This requires that one of two scattering length $a_{\pm}$ can become quite large but the other does not. In three-dimensional free space, $a_{\pm}$ can be obtained by solving the three-body problem with the celebrated Skorniakov-Ter-Martirosian (STM) equation \cite{STM, ren-atom-dimer}. In short, $a_{\pm}$ is related to the $s$-wave channel scattering amplitude $\mathcal{A}^s_{\pm}$ through 
\begin{equation}
a_{\pm}=-\frac{\mu_{{\rm II}, \text{M}}}{2\pi}\mathcal{A}^s_{\pm}\left(k=0, \mathcal{E}=-\frac{\hbar^2}{2\mu_{\rm I,II}a^2_{\rm I,II}}\right),
\label{EqaA}
\end{equation}
where $k$ is the relative momentum between an atom and a molecule, $\mu_{\rm I,II}=m_{\rm I}m_{\rm II}/(m_{\rm I}+m_{\rm II})$ denotes the reduced mass of an A and an AE atom, and $a_{\rm I,II}$ is the scattering length between an A and an AE atom that can be continuously tuned by the magnetic field nearby a Feshbach resonance. 
With the STM equations, the scattering amplitude $\mathcal{A}^s_{\pm}$ is given by 
\begin{align}
\label{stm}
&\frac{\frac{\delta}{(\delta+1)}\mathcal{A}^s_{\pm}(k)}{\frac{1}{a_{\rm I,II}}+
\sqrt{\frac{\delta(\delta+2)}{2(\delta+1)^2}k^{2}+\frac{1}{ a_{\rm I,II}^{2}}}}
\mp\frac{2\pi}{1/a_{\rm I,II}^{2}+k^{2}}
\nonumber\\
&=\pm\frac{(\delta+1)^2}{2\pi(\delta+2)}\int_{0}^{\Lambda} \frac{dk'}{kk'} {\cal F}(k,k') \mathcal{A}^s_{\pm}(k'),
\end{align}
in which ${\cal F}(k,k')$ is defined as 
\begin{equation}
{\cal F}(k,k')\equiv 
\ln\left(\frac{ k^{2}+k'^{2}+\frac{2}{\delta+1}kk'+\frac{1}{ a_{\rm I,II}^{2}}}
{k^{2}+k'^{2}-\frac{2}{\delta+1}kk'+\frac{1}{ a_{\rm I,II}^{2}}}\right).
\end{equation}
Here $\delta=m_{\rm I}/m_{\rm II}$ is the mass ratio, momentum cut-off $\Lambda$ denotes the three-body parameter. By solving Eq. (\ref{stm}) and taking the limit of $k$ goes to zero, one can obtain the atom-molecule scattering length $a_{\pm}$ according to Eq. \eqref{EqaA}.  The results for a mass ratio $m_{\rm I}/m_{\rm II}=1/30$ is shown in Fig. \ref{Fig2} (b).

The results show that $a_{-}/a_{{\rm I,II}}$ is nearly a constant but $a_{+}$ displays a series of resonant behaviors. Thus, nearby the resonance of $a_{+}/a_{{\rm I,II}}$, the spin-exchange interaction proportional to $a_{+}-a_{-}$ will become very large. The underlying reason is because of the famous Efimov states in this three-body system, and the difference in $|+\rangle$ and $|-\rangle$ channels lies in symmetry. Since the $|+\rangle$ channel is symmetric with respect to the exchange of $\uparrow$ and $\downarrow$, it is known that when two out of three scattering lengths are large, there will be Efimov three-body bound states \cite{efimov1, efimov2}. As we schematically shown in Fig. \ref{Fig2} (a), when the energy of one of these Efimov three-body bound states becomes degenerate with the two-body bound state energy, it gives rise to an atom-molecule resonance at which the atom-molecule  scattering length diverges. This is precisely the reason for the resonance of $a_{+}$. On the other hand, it is also known that $s$-wave Efimov states are absent from the antisymmetric channel \cite{petrov}. Indeed, as shown in Fig. \ref{Fig2} (b), there is no resonance of $a_{-}$. Instead, $a_{-}$ exhibits  a linear relation to $a_{\rm I,II}$ with $a_{-}\approx3a_{\rm I,II}$ \cite{petrov}, and $a_{-}$ is also independent on the momentum cutoff $\Lambda$. Hence, we have established the picture that the atom-molecule resonance due to the Efimov three-body bound state can enhance the spin-exchange interaction in this system.

{\it Atom-Molecule Resonance in $3+0$-Mixed Dimensions.} So far we have established that the A-AE atomic mixture nearby Feshbach resonance can be used to simulate the Kondo effect and the Kondo coupling can be enhanced by the atom-molecule resonance originated from the Efimov effect. The later is shown by solving three-body problem in the free space. However, for the Kondo physics, the A atom has to be localized by an external trap. In the rest part of this paper, we will show that the physics of atom-molecule resonance will still hold in the presence of the external trap, although the presence of the trap distorts the discrete scaling symmetry of the Efimov states. 

Now we solve the three-body problem with a harmonic trap for the A atom. The total Hamiltonian is given by 
\begin{align}
&\hat{H}_{0}=-\frac{\hbar^2}{2m_{\rm I}}\nabla_{{\bf r}_{\rm I}}^{2}-\frac{\hbar^2}{2m_{\rm II}}(\nabla_{{\bf r}_{{\rm II},\uparrow}}^{2}+\nabla_{{\bf r}_{{\rm II},\downarrow}}^{2})+\frac{m_{\rm I}}{2}\omega^2r_{\rm I}^{2},
\label{EqH0}\\
&\hat{V}=\sum_{s=\uparrow,\downarrow}\frac{2\pi\hbar^{2} a_{\rm I,II}}{\mu_{\rm I,II}}\delta({\bf r}_{\rm I}-{\bf r}_{{\rm II},s})\frac{\partial}{\partial|{\bf r}_{\rm I}-{\bf r}_{{\rm II},s}|}|{\bf r}_{\rm I}-{\bf r}_{{\rm II},s}|.
\label{EqVbf}
\end{align}
where ${\bf r}_{{\rm I}}$, ${\bf r}_{{\rm II},\uparrow}$ and ${\bf r}_{{\rm II},\downarrow}$ are the coordinates for these three atoms, respectively. To make this problem trackable, we implement the Bohn-Oppenheimer (BO) approximation \cite{efimov2,BOapprox} by considering the mass difference between the two species is sufficiently large (take $^{173}$Yb and $^{7}$Li  for example). Then solving this problem according to the following two steps:

\begin{figure}
\begin{centering}
\includegraphics[width=0.45\textwidth]{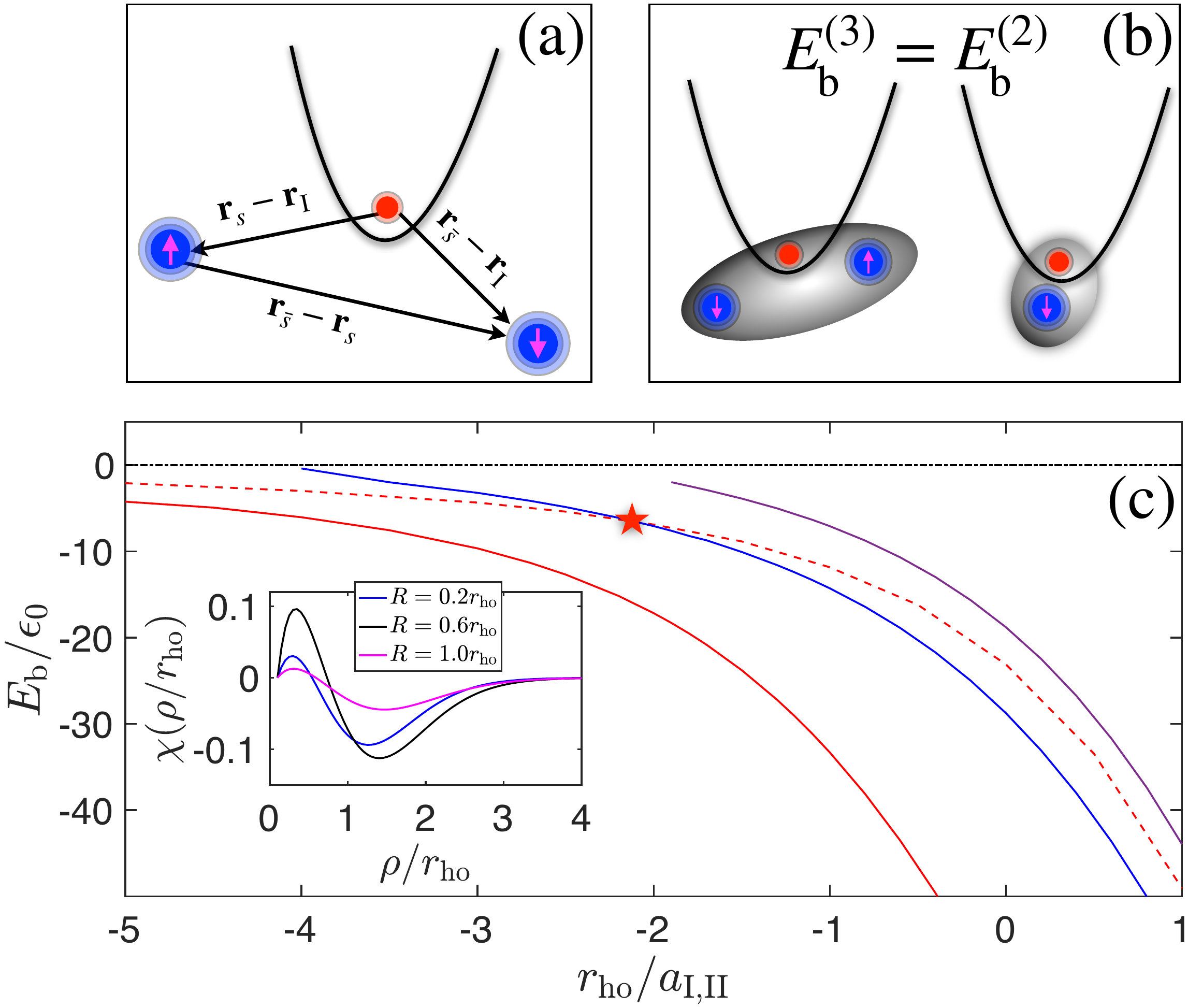}
\caption{(color online) 
(a) Illustration of 3+0 mixed dimensions. The A atom (red ball) is localized by a harmonic trap, and it interacts with two AE atoms (blue balls) with different spins.  
(b) Schematic for the atom-dimer resonance condition in 3+0 mixed dimensions, that is, the binding energy of a trimer formed by one A atom and two AE atoms is degenerate with the binding energy of a dimer formed by one localized A atom and an AE atom. 
(c) Two- and three-body binding states energies are plotted as a function of scattering length $a_{\rm I,II}$. The solid lines are the three-body bound state energy $E_{\rm b}^{(3)}$. The dashed line denotes the two-body bound state energy $E_{\rm b}^{(2)}$. At $r_{\rm ho}/a_{\rm I,II}=-2$ (denoted by the star point), the atom-dimer resonance condition shown in (b) is satisfied. Inset: At the atom-dimer resonance point, the relative wave function $\chi(\rho)=\rho\phi_0(R, \rho)$ for the three-body bound states with various center-of-mass position $R$ is plotted. $\epsilon_0=\hbar\omega/2$ is taken as energy unit and the harmonic trap size $r_{\rm ho}=\sqrt{\hbar/m_{\rm I}\omega}$ is taken as the length unit. The mass ratio $m_{\rm I}/m_{\rm II}$ is fixed at $1/30$. The short range cutoff $\rho_{\Lambda}$ is taken as $0.1r_{\rm ho}$.
}
\label{Fig3}
\end{centering}
\end{figure}

First we fix the positions of two heavy AE atoms, and solve the following Shr\"odinger equation for the A atom as
\begin{equation}
\left[-\frac{\hbar^{2}}{2m_{\rm I}}\nabla^{2}_{{\bf r}_{\rm I}}+\frac{m_{\rm I}}{2}\omega^{2}r_{\rm I}^{2}
+\hat{V}\right]\psi({\bf r}_{\rm I})=E\psi({\bf r}_{\rm I}),
\label{EqBO}
\end{equation}
Generally, the wave function $\psi({\bf r}_{\rm I})$ for Eq. \eqref{EqBO} can be written as a linear combination of two Green's functions \begin{align}
\psi({\bf r}_{\rm I})=\alpha G(E;{\bf r}_{\rm I},{\bf r}_{{\rm II},\uparrow})+\beta G(E;{\bf r}_{\rm I},{\bf r}_{{\rm II},\downarrow}).
\label{EqWF}
\end{align}
Here the Green's function satisfies 
\begin{equation}
\left[E+\frac{\hbar^{2}}{2m_{\rm I}}\nabla^{2}_{{\bf r}_{\rm I}}-\frac{m_{\rm I}}{2}\omega^{2}r_{\rm I}^{2}\right]G(E;{\bf r}_{\rm I},{\bf r}_{{\rm II},s})=\delta({\bf r}_{\rm I}-{\bf r}_{{\rm II},s}),
\end{equation}
where $E$ is the eigenvalue to be determined by the Eq. \eqref{EqBO}. $\alpha$ and $\beta$ are two coefficients to be determined by matching the Bethe-Peierls boundary condition, that is, when ${\bf r}_{\rm I}$ approaches ${\bf r}_{{\rm II},s}$, the wave function $\psi({\bf r}_{\rm I})$ must satisfy 
\begin{align}
\psi({\bf r}_{\rm I}\to{\bf r}_{{\rm II},s})\propto\frac{1}{|{\bf r}_{\rm I}-{\bf r}_{{\rm II},s}|}-\frac{1}{a_{\rm I,II}}.
\label{EqBP}
\end{align}
With this condition, one can obtain $E({\bf r}_{{\rm II},\uparrow}, {\bf r}_{{\rm II}, \downarrow})$ for each set of fixed $\{ {\bf r}_{{\rm II},\uparrow}, {\bf r}_{{\rm II},\downarrow}\}$, which serves as the effective potential for the two AE atoms.

Secondly, this three-body problem is now reduced to a two-body problem with two AE atoms in an effective potential $E({\bf r}_{{\bf II},\uparrow}, {\bf r}_{{\rm II},\downarrow})$. The Schr\"odinger equation is then written as  
\begin{equation}
\left[-\frac{\hbar^2}{2m_{\rm II}}(\nabla_{{\bf r}_{{\rm II },\uparrow}}^{2}+\nabla_{{\bf r}_{{\rm II},\downarrow}}^{2})+E({\bf r}_{{\rm II},\uparrow}, {\bf r}_{{\rm II},\downarrow})\right]\Phi=\mathcal{E}\Phi.
\label{EqHeff}
\end{equation}
In the center-of-mass frame, we write the wave function into different partial wave components as 
$\Phi({\bf R}, \boldsymbol{\rho})=\sum_{l\in \rm even} \phi_l(R, \rho)P_l(\cos\theta) (2l+1)$
where $P_l(x)$ is the Legendre polynomials with orthogonal normalization condition $\int_{-1}^1 dx P_l(x) P_{l'}(x)=\delta_{ll'}2/(2l+1)$, and the $\theta$ is the angle between ${\bf R}=\left({\bf r}_{{\rm II},\uparrow}+{\bf r}_{{\rm II},\downarrow}\right)/2$ and $\boldsymbol{\rho}={\bf r}_{{\rm II},\uparrow}-{\bf r}_{{\rm II},\downarrow}$. Here the wave function only depends on the amplitude of ${\bf R}$ denoted by $R$ because we consider a three-dimensional isotropic trap and there is a rotational symmetry around the center of the trap. It can be found the wave function is dominated by the component with $l=0$, and therefore we only retain the $l=0$ component. We obtain $\phi_0(R, \rho)$ by numerically solving the two-body Shr\"odinger equation Eq. \eqref{EqHeff}.

The calculated three-body bound state energy as a function of $a_{\rm I,II}$ is plotted by solid lines in Fig. \ref{Fig3} (c). A series of three-body bound states emerge with the increasing of $a_{\rm I,II}$ and only the lowest three of them are plotted in Fig. \ref{Fig3} (c).
The corresponding two-body problem of the trapped A atom and another AE atom has been solved by Ref. \cite{yvan}. The energy of one of such two-body bound state is also plotted in Fig. \ref{Fig3} (red dashed line). It can be found that the three-body bound state energy has a cross-point with the two-body bound state energy, which gives rise to same atom-molecule scattering resonance as discussed in the free space case above. In the inset of Fig. \ref{Fig3} (c), we show that the size of three-body bound state is actually comparable to the size of the harmonic trap. 

{\it Conclusion and Outlook.} In summary, we propose a new platform in cold atom system for realizing quantum simulation of the Kondo effect using recently discovered Feshbach resonance between an A and an AE atom. Though for simplicity we only use two nuclear spin components of AE atoms in the above discussion, all discussions apply straightforwardly to $N>2$ nuclear spin components. By combining the advantage of the large nuclear spin of AE atoms and the existence of the Efimov effect in this system, a SU($N$) Kondo model with strong Kondo coupling can be realized here. This already goes beyond condensed matter counter-part of the Kondo physics. Furthermore, using the dynamical control tools in cold atom system, such as a sudden quench of interactions, the highly non-equilibrium dynamics of the SU($N$) Kondo model can also be explored in this system. A successful quantum simulation of the Kondo physics in this setup can certainly enrich our understanding of this problem.

{\it Acknowledgement.} We are indebted to Peng Zhang, Zhenhua Yu, Ran Qi, Pengfei Zhang and Xin Chen for stimulating discussion. Ren Zhang also acknowledges IASTU for the hospitality where part of this work is finished. This work is supported by the National Key R\&D  Program of China Grant No. 2018YFA0307601(RZ) and 2016YFA0301600 (HZ), NSFC Grant No. 11804268 (RZ) and No. 11734010 (HZ).

\end{document}